\documentclass[preprint,showpacs]{revtex4}
\usepackage[centertags]{amsmath}
\usepackage{amsfonts}
\usepackage{amssymb}
\usepackage{amsthm} 
\usepackage{newlfont}
\usepackage{epsfig}
\usepackage{slashbox}
\usepackage{color}
\usepackage{amscd}
\usepackage{graphicx}
\usepackage{amscd}
\usepackage{subfig}

\newcommand{\RR}{{\mathbb R}}
\newcommand{\ZZ}{{\mathbb Z}}

\newcommand{\beq}{\begin{equation}}
\newcommand{\eeq}{\end{equation}}
\newcommand{\ba}{\begin{array}}
\newcommand{\ea}{\end{array}}
\newcommand{\bea}{\begin{eqnarray}}
\newcommand{\eea}{\end{eqnarray}}
\begin{document}

\begin{center}
{\large \sc \bf {Optimization of remote one- and two-qubit state creation via  multi-qubit unitary transformations  at sender and  receiver sides.

}}

\vskip 15pt

{\large 
G.A.~Bochkin and A.I.~Zenchuk 
}

\vskip 8pt

{\it Institute of Problems of Chemical Physics, RAS,
Chernogolovka, Moscow reg., 142432, Russia},\\
% e-mail:  efeldman@icp.ac.ru, zenchuk@itp.ac.ru 

\end{center}

%\today

\begin{abstract}
We study the  optimization problem for   remote one- and two-qubit state creation via a homogeneous spin-1/2 communication line 
using the   local unitary transformations of the multi-qubit sender and  extended receiver.   We show that the maximal length of a communication line used for 
the needed  state creation (the critical length) increases with an increase in the dimensionality of the sender 
and extended receiver.  
The model  with the sender and extended  receiver consisting of up to 10 nodes is used for the one-qubit state creation
and we consider two particular states:  
the almost pure state and  the  maximally mixed one.
Regarding the  two-qubit state creation, we  numerically study the dependence of the critical length on a particular triad of independent  eigenvalues to be created, the model with  four-qubit sender  without an extended receiver is used for this purpose.
 \end{abstract}

\maketitle

%%%%%%%%%%%%%%%%
\section{Introduction}
\label{Sec:Introduction}

The problem of quantum state transfer \cite{Bose} was studied in many papers. The main purpose of that research is  increasing  the 
state-transfer  fidelity in long chains.
Thus,  perfect state transfer is possible in chains with specially adjusted coupling constants governed by  the 
nearest-neighbor XY-Hamiltonian \cite{CDEL,ACDE,KS}, the high probability state transfer (HPST) 
can be arranged in a simpler way using the boundary controlled chains \cite{GKMT,WLKGGB,BACVV2010,ZO,BACVV2011,ABCVV} or  the special non-uniform magnetic field \cite{DZ}. 
The high-fidelity state transfer along the  homogeneous spin chains considered in refs.\cite{H,BOWB} is achieved  via  encoding a
one-qubit state   into the   multi-qubit sender in an optimal way. As an optimization tool,
the singular value decomposition (SVD) of a certain matrix was used.

In this paper  we consider  the remote one-qubit  state creation  in a homogeneous  communication line  with 
a multi-qubit  sender. The  long distance creation of a needed state  is achieved using a
pair of optimized local unitary transformations on the sender and receiver sides. This  optimization is based on the SVD of a certain matrix 
whose elements are expressed in terms of transition amplitudes between different nodes of a communication line.

Among the one-qubit creatable states, we consider the  almost pure state (in this case we deal with an analogue of 
high-probability state transfer)
and the maximally mixed state (i.e., the  state with two equal eigenvalues $\lambda_1=\lambda_2=\frac{1}{2}$). The interest in the latter state is motivated by the fact \cite{BZ_2015} that 
having the communication line allowing  the creation of  maximally mixed state we can also create the state with 
any eigenvalue $\lambda_1$ just using the proper choice of the 
parameters of  the sender's initial state.  Hereafter the maximal length allowing the particular state creation is referred to as the critical length $N_c$ for this state. Thus, the  critical length for the  
maximally mixed state  is also the critical length for the creation of a state with an arbitrary eigenvalue.
 
We also consider the remote two-qubit state creation. Since this state is multi-parametric one, we restrict ourselves to 
studying the eigenvalue creation disregarding other parameters of that state. At that, we numerically find the 
dependence of the critical length $N_c$ on the eigenvalues to be created.

The paper is organized as follows. The general protocol of remote state creation using the communication line with multi-qubit 
sender and extended receiver is given in Sec.\ref{Section:phase}. The optimization is based on SVD. This protocol is applied to the 
one-qubit state creation in Sec.\ref{Section:1qubit} where the communication line with the sender and extended receiver of 
up to 10 nodes is used. We consider  the high-probability almost pure  state creation and creation of maximally mixed state.
The eigenvalue creation of two-qubit receiver is studied in Sec.\ref{Section:2qubit}. Conclusions are given in Sec.\ref{Section:conclusion}.
The time optimization and some details of two-qubit state creation are given in Appendix, Sec.\ref{Section:Appendix}.

%The problem of controllable remote state creation is considered in set of papers 
%\cite{PBGWK2,PBGWK,DLMRKBPVZBW,XLYG,PSB}, and it 
%initiates  the 
%problem of remote creation of states with desirable quantum correlations. 
%In particular, the  entanglement between the remote qubits is studied in  \cite{BBVB,CS}, 
% different method of creating the quantum correlations are considered in 
% \cite{DSC,LBAW,NLLZ,ZC,SXSZDWHCKW,RDL}.
 
\section{One-excitation spin dynamics and optimization tool }
\label{Section:phase}
\begin{figure*}
   \epsfig{file=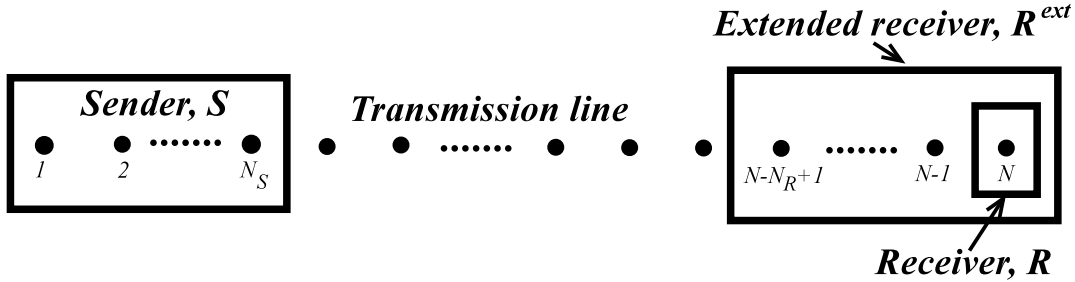,
  scale=0.5
   ,angle=0%270
}
\caption{The communication line with the $N_S$-qubit sender $S$, one-qubit receiver $R$ and $N_R$-qubit extended receiver $R^{ext}$.
} 
  \label{Fig:SR} 
\end{figure*}

We proceed  with the one-qubit receiver and consider  the problem of high-probability pure state creation \cite{KZ_2008,SAOZ,BK,C,QWL,B,JKSS,SO,BB,SJBB} and eigenvalue creation \cite{BZ_2015,BZ_2016} 
in a long spin-1/2 chain with the $N_S$-qubit sender, one-qubit receiver and $N_R$-qubit extended receiver, as  shown in Fig.\ref{Fig:SR}.

Hereafter  we consider the one-excitation spin dynamics governed by 
the Hamiltonian $H$ conserving the  $z$-projection of the total spin. 
%In addition, it is shown in  ref.\cite{BZ_2015} that the diagonal  one-qubit 
%creatable states correspond to the completely polarized  initial state, i.e., the initial state is a superposition of one-excitation sender's 
%states  whithout a contribution from the ground state.
{The evolution of a one-excitation state  can be described in the $N$-dimensional 
space spanned by the following basis vectors:
\begin{eqnarray}\label{bas}
|n\rangle, \;\;n=1,\dots,N,
\end{eqnarray}
where $n$ means the state with the $n$th polarized spin. 
%The critical length of a particular state is such a length of communication line that this state belongs to the boundary. 
%In this paper we calculate only the critical lengths, so that we deal with the spin dynamics in the basis (\ref{bas}).
Therefore, the general initial state $|\Psi_0\rangle$  of our interest  reads:}
\begin{eqnarray}\label{InitialSt}
|\Psi_0\rangle  =\sum_{i=1}^{N_s} a_i | i\rangle, \;\;\sum_{i=1}^{N_s} |a_i|^2=1 .
\end{eqnarray}
Its evolution  is described by   the Lioville equation:
\begin{eqnarray}\label{ev}
|\Psi(t)\rangle = e^{-i H_1 t} |\Psi_0\rangle,
\end{eqnarray}
where $H_1$ is the $N\times N$ block   of the Hamiltonian responsible for the one-excitation evolution. We  
  can diagonalize the Hamiltonian $H_1$:
 \begin{eqnarray}\label{Hd}
 H_1=W e^{-i \Lambda t} W^+,
\end{eqnarray}
where $\Lambda ={\mbox{diag}}(\lambda_1,\dots,\lambda_N)$ and $W$  are, respectively, the eigenvalue matrix and the matrix of the eigenvectors.
If, in addition, we apply the unitary transformation $V$ to the extended receiver at the time instant $t_0$, then 
the obtained state $\Psi^V$ reads
\begin{eqnarray}\label{evV}
|\Psi^V\rangle = \tilde V  W e^{-i \Lambda t_0} W^+ |\Psi_0\rangle.
\end{eqnarray}
Here the operator $\tilde V$ in the basis (\ref{bas}) has the  block-diagonal 
form, $\tilde V={\mbox{diag}}(I_{rest},V)$, where
$ I_{rest}$ is an $(N-N_R)$-dimensional identity operator 
and $V$ is  an $N_R\times N_R$ unitary operator.  %This form of the  transformation $\tilde V$ is proposed by the result of ref. \cite{BZ_2015} that 
%the upper boundary of the remotely creatable region corresponds to $a_0=0$, and other  states (inside of the creatable region)  correspond to $a_0>0$. Therefore we take $\tilde V|0\rangle =|0\rangle$ that is used in obtaining  formula (\ref{evV}). 
The state of the one-qubit receiver $\rho^R$ is determined by the trace of the state (\ref{ev}) over all other spins:
\begin{eqnarray}\label{RhoR}
\rho^R = {\mbox{Tr}}_{1,\dots,N-1} |\Psi^V\rangle \langle \Psi^V|,
\end{eqnarray}
where the trace is taken over all the nodes except the node of the receiver.
It is simple to show (see, for instance,  ref.\cite{BZ_2015})
that this state has the following diagonal form (represented in the one-qubit basis $|0\rangle$, $|1\rangle$)
\begin{eqnarray}\label{R}
\rho^R = \left(
\begin{array}{cc}
1-|f_N|^2 & 0\cr
0 & |f_N|^2 
\end{array}
\right),
\end{eqnarray}
where the projection $f_N$ is defined  as
\begin{eqnarray}\label{fj}
f_N=\langle N|\Psi^V\rangle,
%,\;\;n=1,\dots,N.
\end{eqnarray}
and  the eigenvalues of state (\ref{R}) read
\begin{eqnarray}
%\lambda_1=\frac{1}{2}\left(1+\sqrt{(1-2 R_N^2)^2+ 4 R_N^2 R_0^2}\right),\;\;\lambda_2=1-\lambda_1.
\lambda_1=1-|f_N|^2,\;\;\lambda_2=|f_N|^2.
%\frac{1}{2}\left(1+\sqrt{(1-2 |f_N|^2)^2+ 4 |f_N|^2 |f_0|^2}\right),\;\;\lambda_2=1-\lambda_1.
\end{eqnarray}
Below we consider the creation of two particular states. The first one is the state maximally approximating the pure state
$|N\rangle$ (an analogy of the 
HPST). More exactly, the state with
\begin{eqnarray}\label{HPST}
|f_N|^2 > 0.9.
\end{eqnarray}
The second state is the maximally mixed state, $\lambda_1=\lambda_2=\frac{1}{2}$, which corresponds to 
\begin{eqnarray}\label{mixed}
|f_N|^2 = \frac{1}{2}.
\end{eqnarray}
{The motivation for studying this state is mentioned in the Introduction. 
The choice of the above two states and appropriate conditions (\ref{HPST}) and (\ref{mixed}) support our
disregarding the contribution to the initial state  (\ref{InitialSt}) from the ground state  (i.e., the state without excitations), because this contribution reduces $|f_N|$ \cite{BZ_2015}.} 
%The interest to this state is motivated by the fact that the communication line creating the one-qubit state with two equal eigenvalues 
%can create a  state with an arbitrary pair of eigenvalues (constrained by the relation $\lambda_1+\lambda_2=1$) by varying the parameters of the %initial sender's state \cite{BZ_2015}.   

%\newpage

%%%%%%%%
\subsection{Singular-value decomposition as optimization tool}
\label{Section:SVD}
For constructing the desired state we use the optimization method based on the SVD \cite{H} (see also \cite{BZ_2016}).
We give some details of this procedure.
%First of all, we note that the maximum of $|f_N|$ corresponds to  $a_0=0$ in 
%(\ref{InitialSt}) \cite{BZ_2015}. Therefore
The projection  $f_N$ defined in  (\ref{fj}) can be represented as follows:
\begin{eqnarray}\label{fN}
&&f_N= \langle N|\Pi_R^+ V {\cal{P}} \Pi_S| \Psi_0 \rangle, \;\; 
{\cal{P}}_{nm} =  \sum_{k=1}^N W_{(N-N_R+n) k} e^{-i \lambda^H_k t} W_{mk}, \\\nonumber
&&n=1,\dots,N_R, \;\;m=1,\dots,N_S,
\end{eqnarray}
where we take into account reality of $W$ and use two shorten  bases: 
$|n\rangle_S$, $n=1,\dots,N_S$ (to enumerate the columns  of ${\cal{P}}$)  
and $|n\rangle_R$, $n=1,\dots,N_R$ 
(to enumerate the rows  of ${\cal{P}}$) introduced via the rectangular operators  
$\Pi_S$ ($N_S\times N$)  and  $\Pi_R$ ($N_R\times N$) %on the subspaces of, respectively, the  sender and the extended receiver:
by the formulas
\begin{eqnarray}\label{SRbasis}
&&
|n\rangle_S =\Pi_S |n\rangle,\;\;n=1,\dots,N_S, \;\;\Pi_S |n\rangle =|0\rangle_S,\;\;n>N_S,\\\nonumber
&&
|n\rangle_R =\Pi_R |N-N_R+n\rangle,\;\;n=1,\dots,N_R,\;\;\Pi_R |n\rangle =|0\rangle_R,\;\;n\le N-N_R.
\end{eqnarray}
Thus, $\Pi_S$ and $\Pi_R$ are, respectively, the first $N_S$ and the last $N_R$ rows of the $N\times N$ identity matrix. 
SVD  of ${\cal{P}}$  reads
\begin{eqnarray}\label{SVD}
{\cal{P}} =V^{SVD} \Lambda^{SVD} (U^{SVD})^+,
\end{eqnarray}
where $\Lambda^{SVD}={\mbox{diag}} ( \omega_1,\omega_2\dots)$ is the diagonal matrix of singular values. We require that 
 the first
singular value is the maximal one. 
We note two symmetry properties of SVD.
\begin{enumerate}
\item
 SVD is defined up to the
phase transformation
\begin{eqnarray}\label{SVDsym1}
(V^{SVD}_{kn}, U^{SVD}_{mn}) \;\to \; (V^{SVD}_{kn} e^{i \varphi_n}, U^{SVD}_{mn} e^{i \varphi_n}),\;\;\forall \varphi_n\in \RR.
\end{eqnarray}
\item
The considered system is symmetrical with respect to the reversion of the order of the nodes 
 (the Hamiltonian is symmetrical with respect to the secondary diagonal). Therefore, if we consider the  chain  with 
 $N_S=N_1$, $N_R=N_2$ (${\cal{P}}_1=V^{SVD}_1\Lambda^{SVD}_1 (U^{SVD}_1)^+$)
and the other chain with    $N_S=N_2$, $N_R=N_1$ (${\cal{P}}_2=V^{SVD}_2\Lambda^{SVD}_2 (U^{SVD}_2)^+$), then (up to the above phase transformation)
\begin{eqnarray}\label{SVDsym2}
\Lambda^{SVD}_2 = (\Lambda^{SVD}_1)^T, \;\; (V^{SVD}_2)_{n,m} = (U^{SVD}_1)^*_{N_1-n+1,m},\;\;  (V^{SVD}_1)_{n,m}= (U^{SVD}_2)^*_{N_1-n+1,m}.
\end{eqnarray}
In particular, if $N_1=N_2$, then both chains are equivalent and
\begin{eqnarray}\label{SVDsym22}
\Lambda^{SVD}_1 \equiv \Lambda^{SVD}_2, \;\; (V^{SVD}_1)_{n,m}= (V^{SVD}_2)_{n,m} = (U^{SVD}_1)^*_{N_1-n+1,m} = (U^{SVD}_2)^*_{N_1-n+1,m},
\end{eqnarray}
where $*$ means complex conjugation.
\end{enumerate}
Next, we  introduce the $N_S\times N_S$ unitary operator $U$ of the sender such that $\Pi_S| \Psi_0 \rangle = U \Pi_S|1\rangle$ and rewrite eq.(\ref{fN}) as
\begin{eqnarray}
\label{fN2}
f_N=~\langle N|\Pi_R^+ V V^{SVD}\Lambda^{SVD}(U^{SVD})^+ U \Pi_S|1\rangle.
\end{eqnarray}
Remark that the operator  ${\cal{P}}$ (\ref{SVD}) is independent on the  parameters $a_i$ 
and is completely defined by the Hamiltonian. To maximize $|f_N|$, 
we require that 
\begin{eqnarray}\label{U2}
&& U\Pi_S|1\rangle=U^{SVD}\Pi_S|1\rangle,\\\label{V}
&& \langle N| \Pi_R^+ V=\langle 1| \Pi_R^+ (V^{SVD})^+,
\end{eqnarray}
which hold if the operators $U$ 
and $V$  satisfy the following equations:
\begin{eqnarray}
\label{U}
U=U^{SVD},\;\;\;%\sum_{j=1}^{N_R} V_{N_Rj}V^{SVD}_{jk} =\delta_{k1}.
\sum_{j=1}^{N_R} V_{nj}V^{SVD}_{jk} =\delta_{N_R-n+1,k}.
\end{eqnarray}
Now, substituting expressions (\ref{U2}) and (\ref{V}) into  eq.(\ref{fN2}) we finally obtain
\begin{eqnarray}
f_N=w_1.
\end{eqnarray}
Note, that the column-vector $U\Pi_S|1\rangle$ (\ref{U2})  and the row-vector 
$\langle N|\Pi_R^+ V$ (\ref{V}) in eq.(\ref{fN2}) are defined, respectively, by the first column of the matrix $U$ 
(or $U^{SVD}$) and by
 the $N_R$th row of the matrix $V$ (the first column of the matrix  $V^{SV}$). Vectors (\ref{U2}) and (\ref{V}) will be used as characteristics of the optimization protocol in Sec.\ref{Section:1qubit}.

%The 
%elements $U^{SVD}_{1k}$ and $V^{SVD}_{k1}$ are shown in  Fig.\ref{} 

\subsection{Spectral analysis}
To clarify the mechanism of obtaining the desired state we turn to the  
 spectral representation of the projection $f_N$: 
\begin{eqnarray}\label{fNs}
&&
f_N=\langle N|\Psi^V(t)\rangle  = \langle N| \tilde V W e^{-i \Lambda t} W^+ |\Psi_0\rangle=\\\nonumber
&&
\sum_k \sum_{m=1}^{N_R} V_{N_Rm} W_{(N-N_R+m) k} e^{-i \lambda^H_k t} \sum_{j=1}^{N_S} W_{jk} a_j  = \sum_k P_{Nk}  e^{-i \lambda^H_k t + i \phi_{Nk}},
\end{eqnarray}
where the spectral amplitudes  $P_{Nk}$ and the optimizing spectral phases $\phi_{Nk}$ read 
\begin{eqnarray}\label{P}
 P_{Nk} = |\sum_{m=1}^{N_R} V_{N_Rm} W_{(N-N_R+m) k}  \sum_{j=0}^{N_S} W_{jk} a_j|,\;\;  
 \phi_{Nk} =Arg(\sum_{m=1}^{N_R} V_{N_Rm} W_{(N-N_R+m) k}  \sum_{j=0}^{N_S} W_{jk} a_j).
\end{eqnarray}
 Eq.(\ref{P}) shows that, for a given spectrum $P_{Nk}$ (which depends on the  parameters of the initial state $a_i$, 
 on the local unitary transformation $V$ and on the Hamiltonian $H_1$), the  optimization must be aimed on creating 
 such phases $\phi_{ik}$ that, at the optimal time instant $t_0$,
we have, in the ideal case,
\begin{eqnarray}\label{phases}
  \Phi_{Nk}\equiv  \phi_{Nk}-\lambda^H_k t_0=2 \pi n_k, \;\;\forall \;k, \;\; n_k\in \ZZ.
\end{eqnarray}
Then all harmonics $P_{Nk}$ would maximally contribute  to $|f_N|$. 
Of course, this ideal case is realizable only in the case of   perfect state transfer. 
In general, having the  unitary transformations $U$ and $V$  of, respectively, the $N_S$-dimensional  sender and 
 the $N_R$-dimensional extended receiver
we can not adjust all $N$ phases. But we can adjust those of them which correspond to the valuable spectral amplitudes $P_{Nk}$,
\begin{eqnarray}\label{Pmin}
P_{Nk} > P_{min},
\end{eqnarray}
where $P_{min}$ is some conventional parameter.
In other words, maximizing $|f_N|$, we have to 
provide large values only for those $P_{Nk}$, for which 
\begin{eqnarray}\label{phasesApprox}
\phi_{Nk}-\lambda^H_k t_0\approx 2 \pi n_k.
\end{eqnarray}
Of course, conditions (\ref{Pmin},\ref{phasesApprox}), in general, can be satisfied only over some spectral interval,
\begin{eqnarray}\label{kint}
k_{min} <k < k_{max}.
\end{eqnarray}
%and the spectral amplitudes   $P_{Nk}$  inside of the interval (\ref{kint}) :

%Then, to obtain the desired result of the state transfer/creation we have to take care only about the phases $\phi_{Nk}$  with $k$, 
%corresponding to the interval of eigenvalues  (\ref{kint}).

\subsection{XY-Hamiltonian}

Below we consider the spin dynamics in a  particular model of  communication line governed by the $XY$-Hamiltonian with all-node interaction:
%We consider the spin dynamics, governed by the $XY$-Hamiltonian with all-node interactions 
\begin{eqnarray}\label{XY}
H=\sum_{j>i} D_{ij}(I_{ix}I_{jx}+ I_{iy}I_{jy}),\;\;D_{ij}=\frac{\gamma^2 \hbar}{r_{ij}^3},
\end{eqnarray}
where $\hbar$ is the Planck constant,   $\gamma$ is the gyromagnetic ratio, $r_{ij}$ is the distance between the $i$th and the $j$th spins, $I_{ i\alpha}$
$(\alpha = x, y, z)$ is the projection operator of the $i$th spin on the $\alpha$ axis, $D_{ij}$ is the dipole-dipole
coupling constant between the $i$th and the $j$th nodes. 
Below we use the dimensionless time
assuming $D_{12} = 1$.
In all numerical experiments described below, the  optimization yields the  time instant $t_0\approx N$.

\section{One-qubit state creation}
\label{Section:1qubit}

\subsection{High probability creation of excited one-qubit state}

In this section we use the protocol of Sec.\ref{Section:SVD} to optimize the communication line for the purpose of 
high probability pure state $|N\rangle$ creation
(see condition (\ref{HPST}))
in a long communication line.
Optimizing the singular value $w_1$ in SVD (\ref{SVD}) for the communication lines  we find the critical  length 
$N_c$ for this state creation for different dimensionalities 
$N_S$ and $N_R$.
 The results of our calculations are collected  in Table \ref{Table:NSNRaHPST}.

As we noticed in Sec.\ref{Section:phase}, the state creation can be characterized by the spectral amplitudes  $P_{Nk}$ and  phases $\Phi_{Nk}$. We show that, after the  optimization of $w_1$,  
 condition (\ref{phases}) is approximately satisfied for those $k$ for which the spectral amplitudes $P_{Nk}$ are essentially bigger then zero. 

\begin{figure*}
\subfloat[]{\includegraphics[scale=0.4,angle=0]{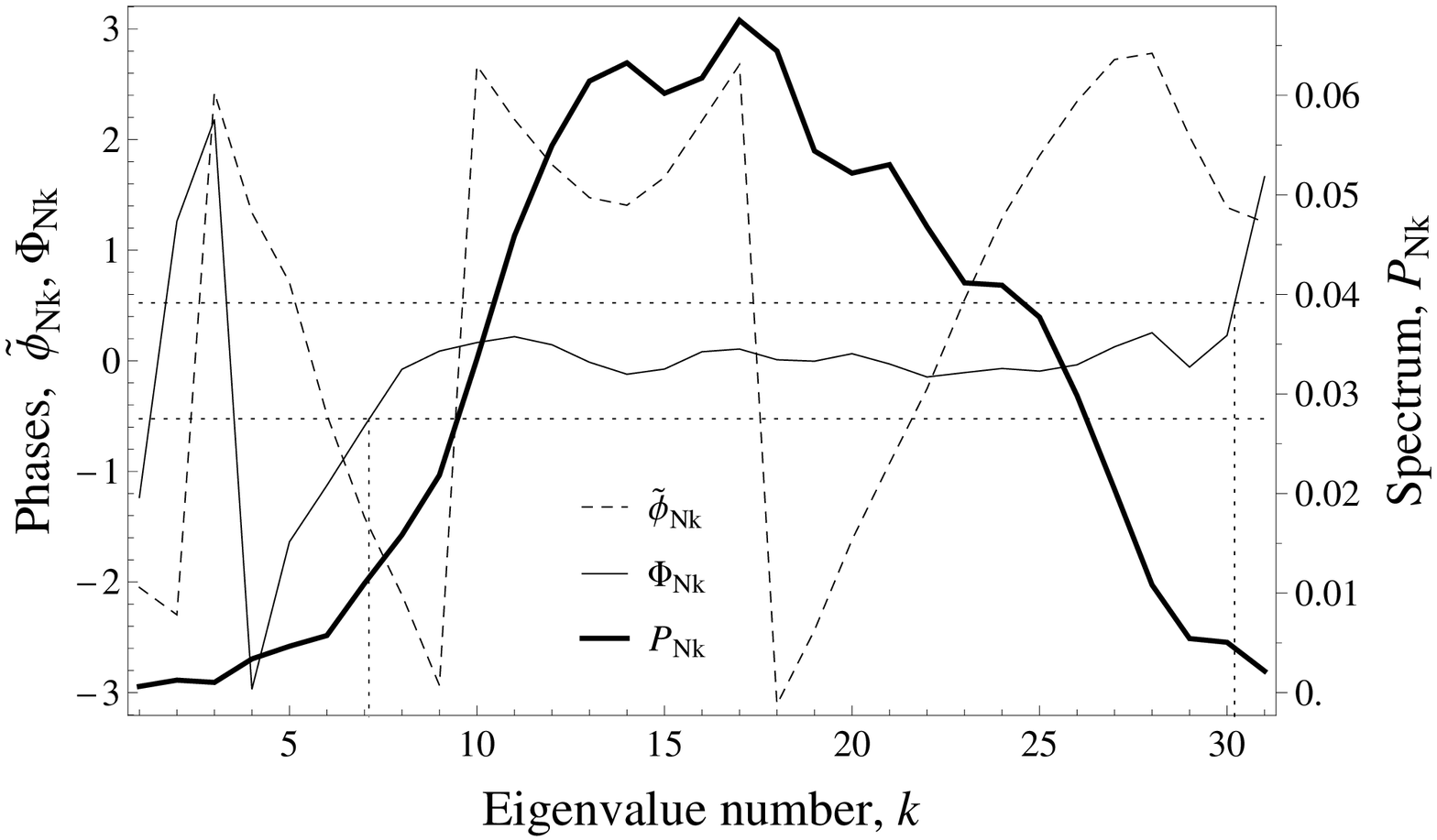}}
\subfloat[]{\includegraphics[scale=0.4,angle=0]{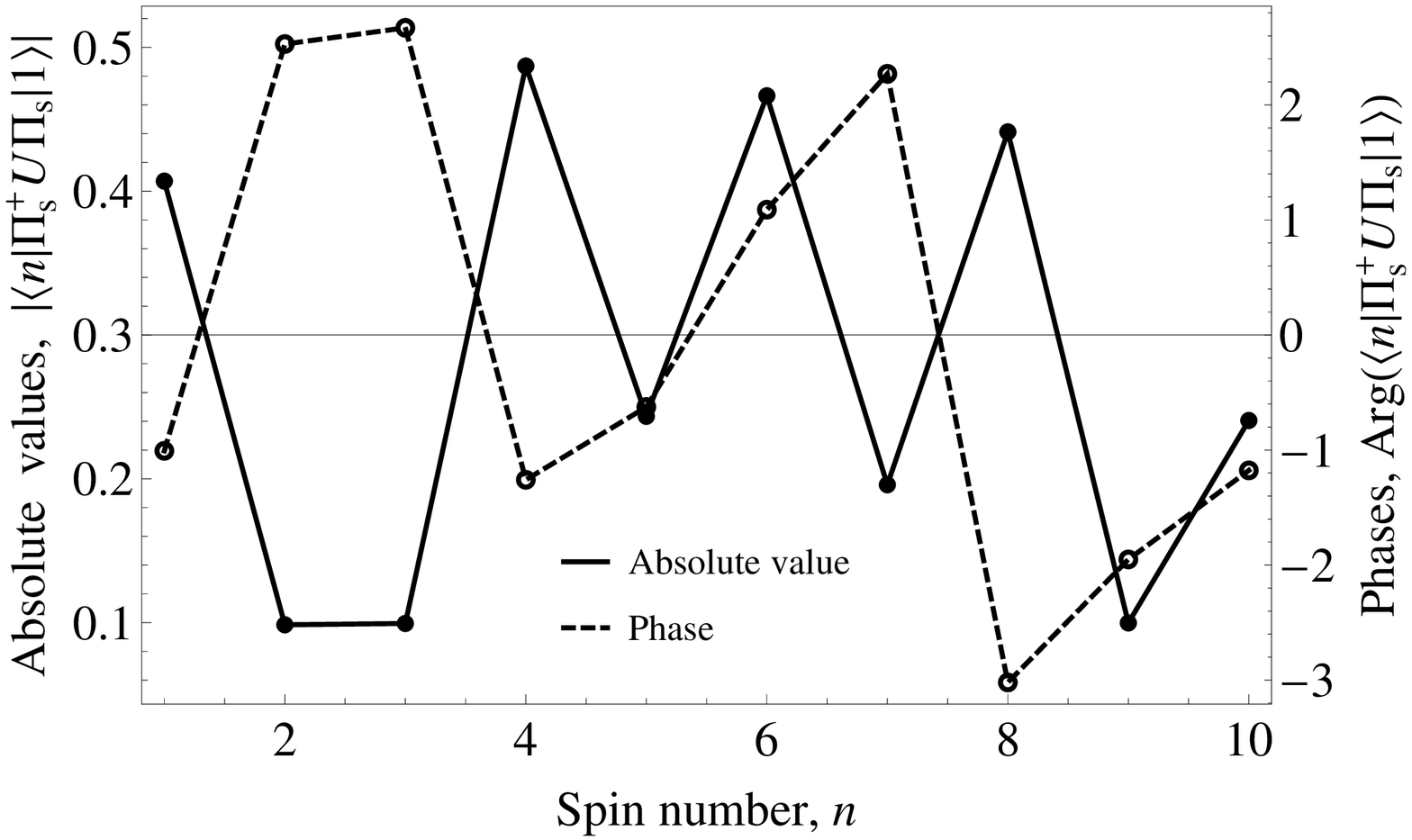}}\\
\subfloat[]{\includegraphics[scale=0.4,angle=0]{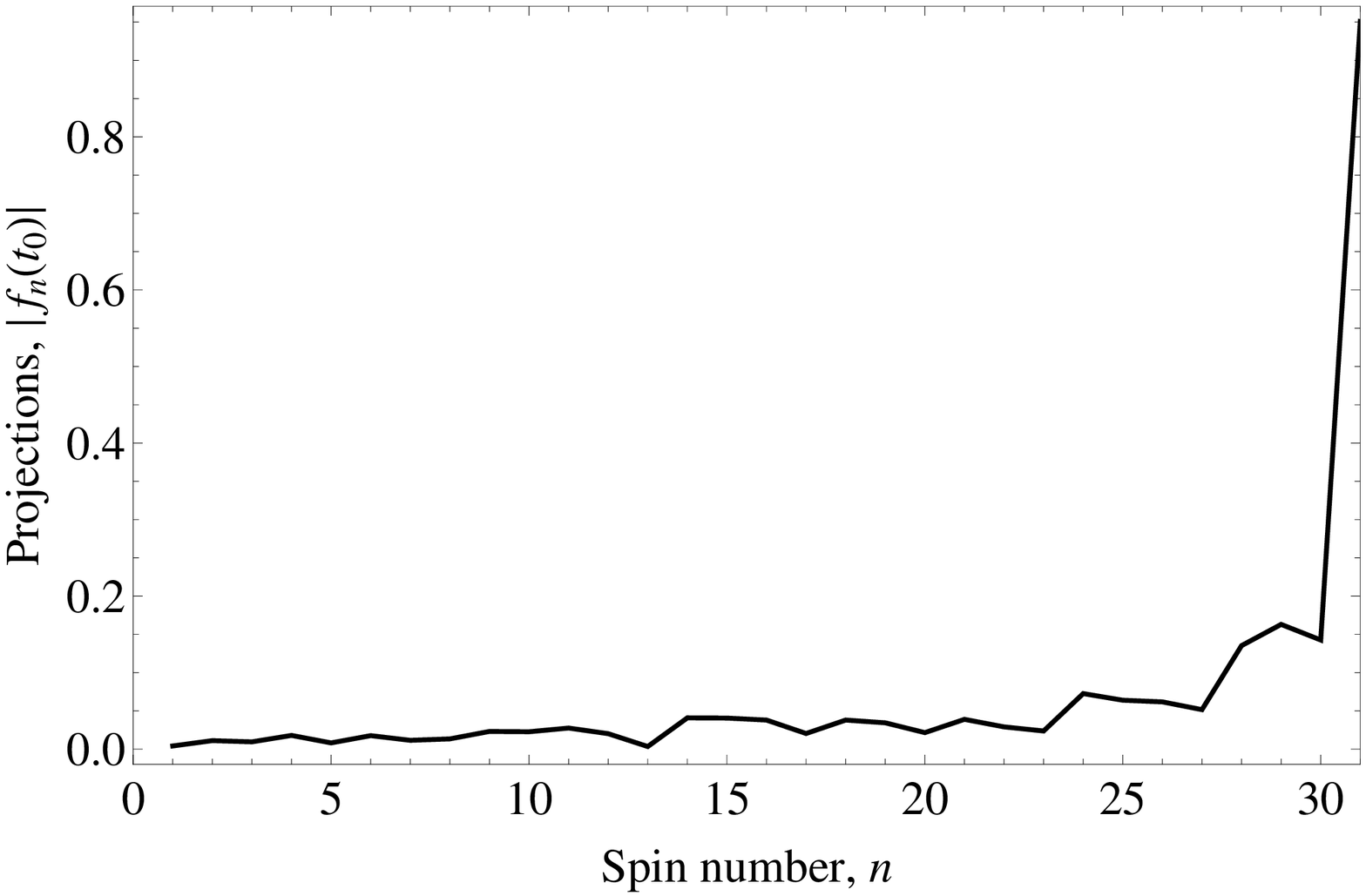}}
\caption{
The HPST (\ref{HPST}) in communication line of $N_c=31$ nodes with   $N_S=10$ and $N_R=1$ at $t_0=39.3815$.
(a) The spectrum $P_{Ni}$, the optimizing phases $\tilde \phi_{Nk}=\phi_{Nk} + \pi k$ and the 
resulting phases $ \Phi_{Nk}= \phi_{Nk}  - \lambda^H_{k} t_0$;
(b) The amplitudes and the phases of the elements of the vector (\ref{U2}) (or the initial projections 
$f_{n}(0)=\langle n|\Pi_S^+ U \Pi_S|1\rangle$);
(c) The profile of the absolute values of the  projections $|f_{n}(t_0)|$.
} 
  \label{Fig:P1ExtR1} 
\end{figure*}

For instance, we consider  the communication line with $N_S=10$  without the  extended receiver (i.e., $N_R=1$). In this case $N_c=31$, 
the corresponding spectrum is shown in Fig.\ref{Fig:P1ExtR1}a (thick solid line). The optimizing phases $\phi_{Nk}$
and the resulting phases $\Phi_{Nk}= \phi_{Nk}  - \lambda^H_{k} t$     are shown in the same figure (see, respectively, the dashed and thin-solid lines;
for convenience of visualization, we show $\tilde \phi_{Nk}=\phi_{Nk} +\pi k$ instead of $\phi_{Nk}$).
We conclude  that almost the  whole spectrum is involved in the state-transfer process (almost 
all possible $P_{N_k}$ are essentially non-zero), and almost all 
these  harmonics appear with the proper phase. In fact,  assuming that the condition (\ref{phasesApprox}) is satisfied if 
spread of phases is inside of  the   interval (see the horizontal dotted lines in Fig.\ref{Fig:P1ExtR1}a):
\begin{eqnarray}\label{PhiInt}
-\frac{\pi}{6} < \Phi_{Nk} < \frac{\pi}{6},
\end{eqnarray}
we find  the corresponding spectral interval 
 $8 \le k \le 30$ (the vertical dotted lines in Fig.\ref{Fig:P1ExtR1}a), which involves almost all harmonics. 

\begin{figure*}
\subfloat[]{\includegraphics[scale=0.4,angle=0]{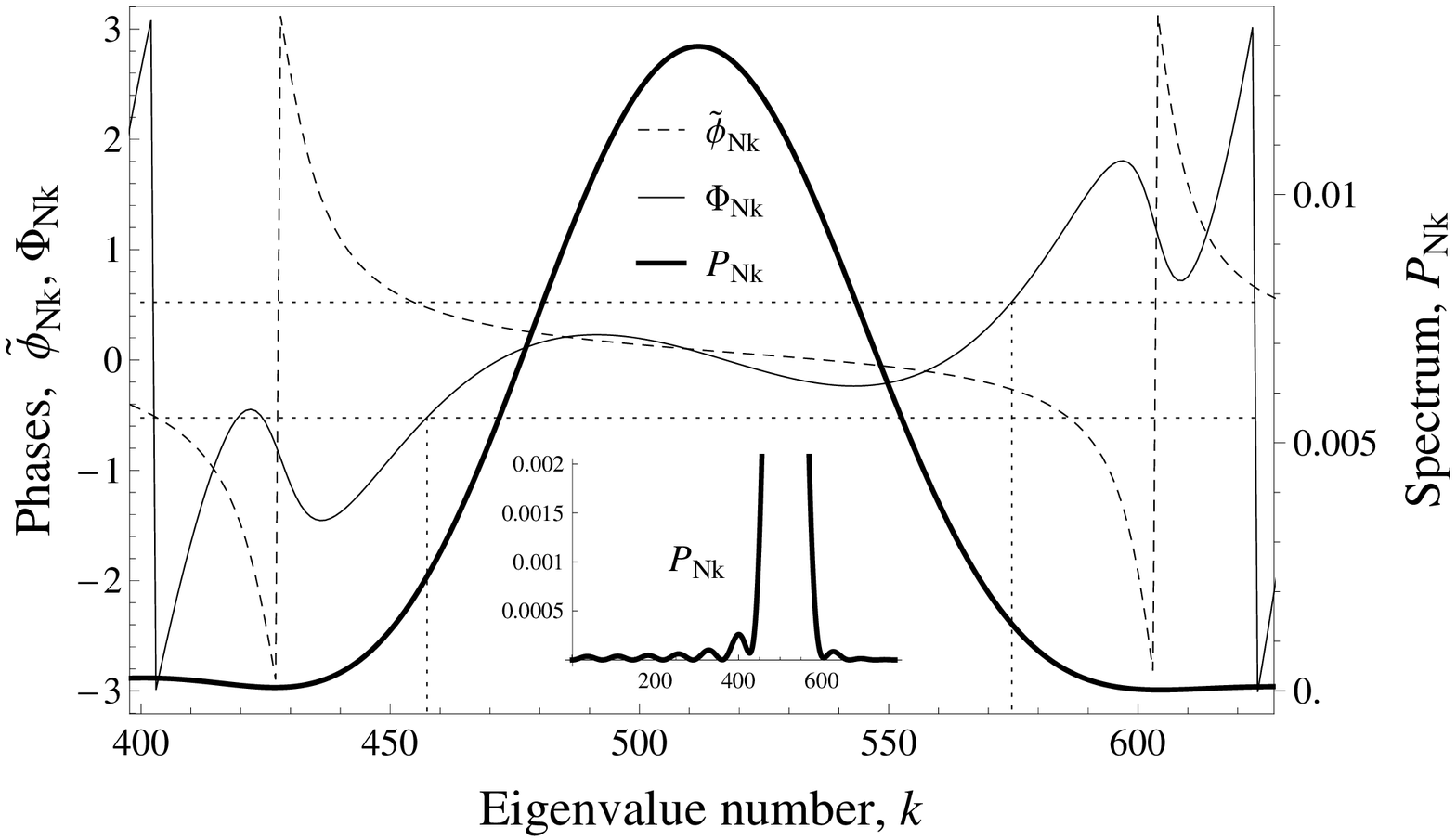}}
\subfloat[]{\includegraphics[scale=0.4,angle=0]{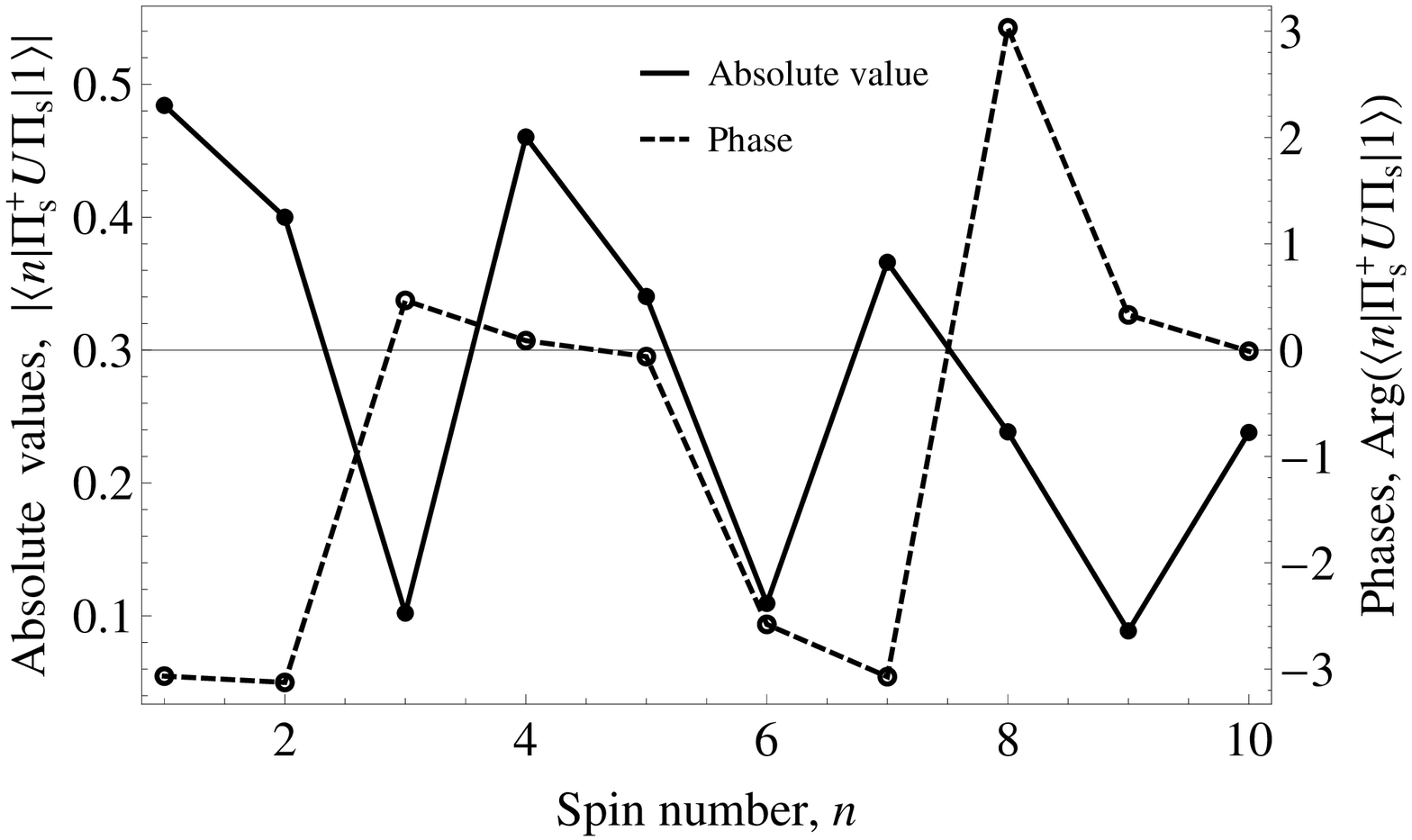}}\\
\hspace*{-0.95cm}\subfloat[]{\includegraphics[scale=0.47,angle=0]{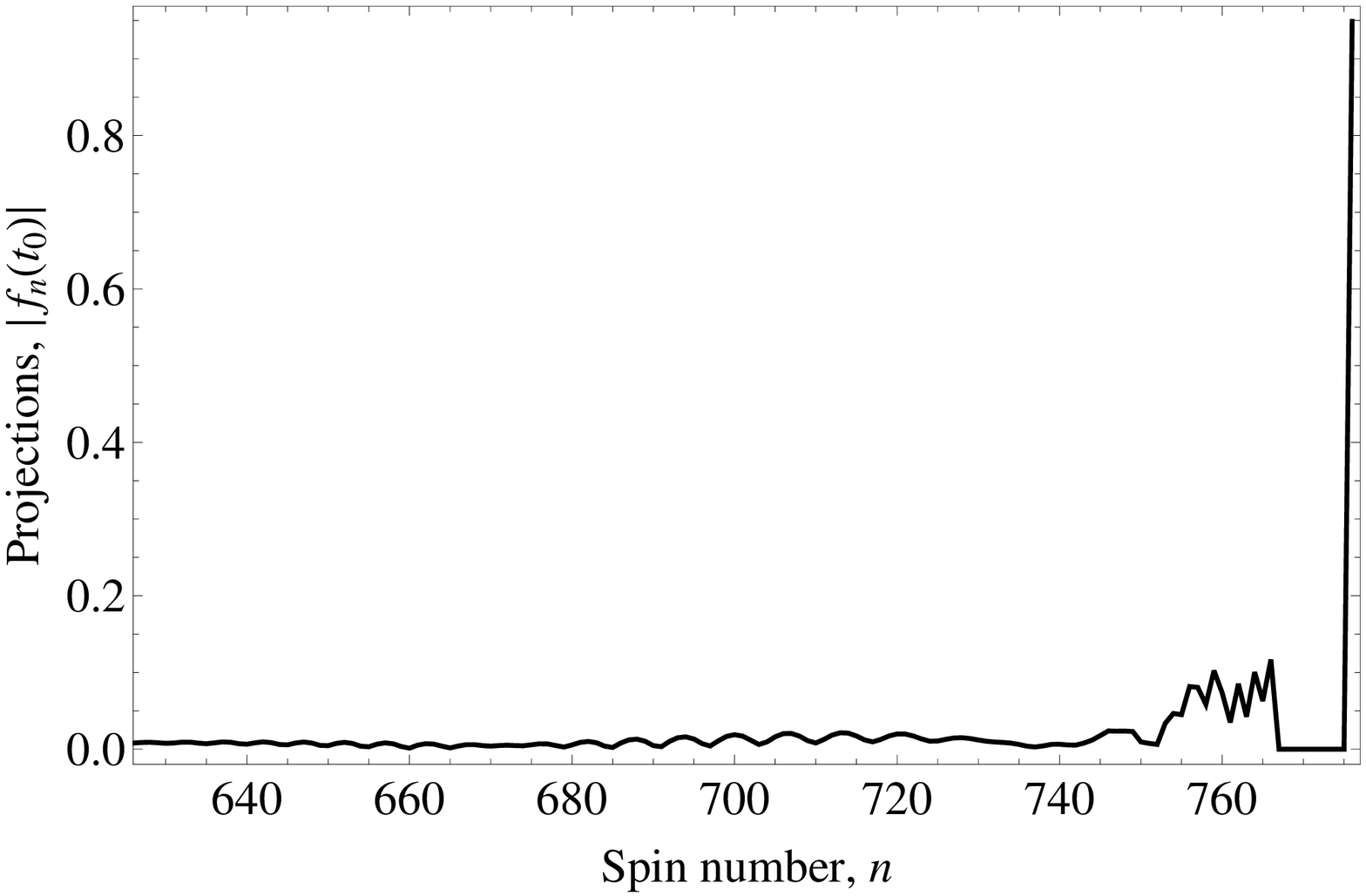}}

\vspace*{-7.9cm}

\subfloat{\includegraphics[scale=0.4,angle=0]{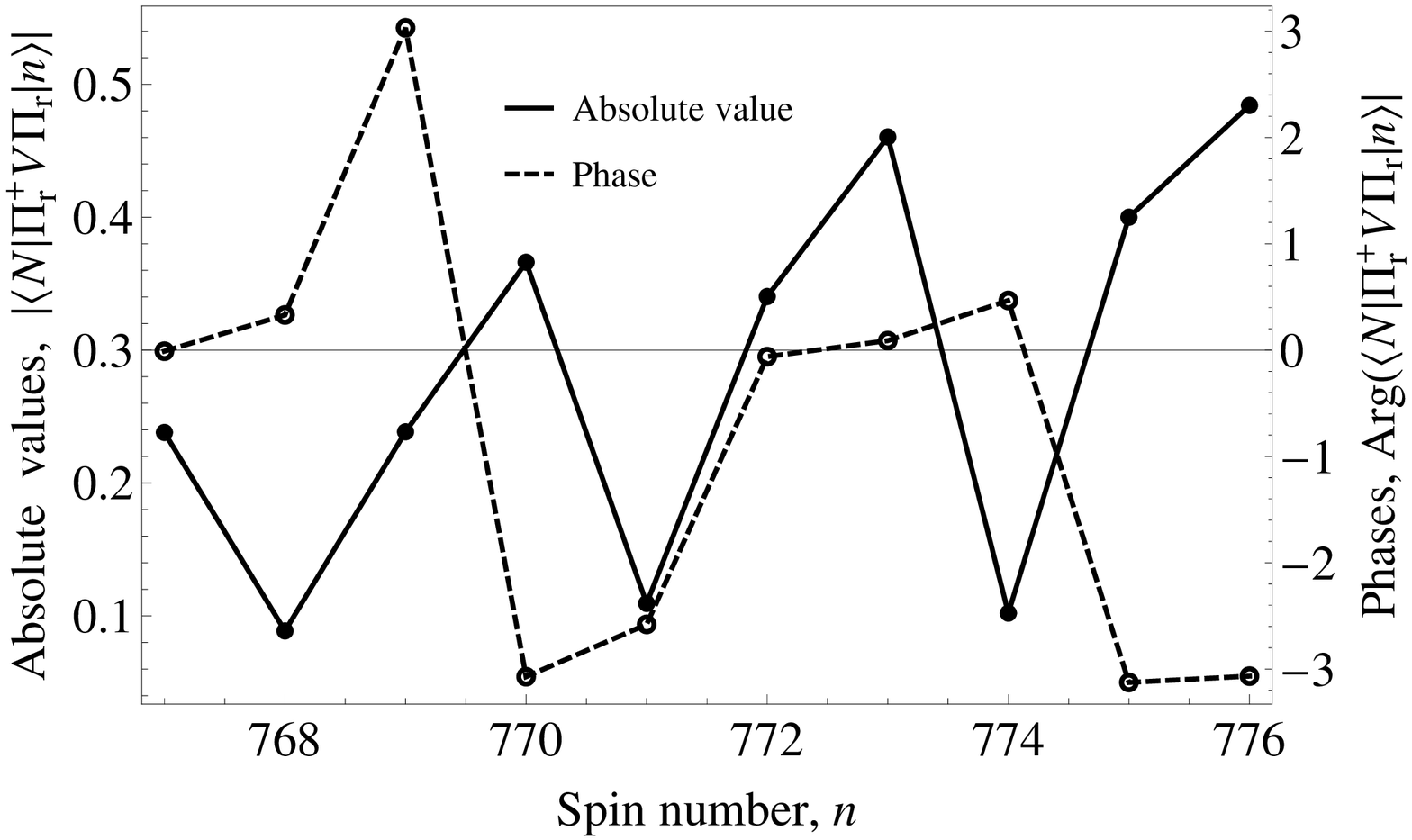}}

\vspace*{2cm}

\caption{%$R\sim 1$ n 776 rsize 10, t=767.892
The HPST (\ref{HPST}) in communication line of $N_c=776$ nodes with   $N_S=N_R=10$  at $t_0=767.892$.
(a) The spectrum $P_{Ni}$, the optimizing phases $\tilde \phi_{Nk}=\phi_{Nk} +\pi k$ and the 
resulting phases $ \Phi_{Nk}= \phi_{Nk}  - \lambda^H_{k} t_0$. The inset represent the spectrum  $P_{Nk}$ over the whole spectral interval;
(b) The  amplitudes and the  phases of the elements of the vector (\ref{U2}) (or the initial projections 
$f_{n}(0)=\langle n|\Pi_S^+ U \Pi_S|1\rangle$);
(c) The profile of  the absolute values of the 
projections  $|f_{n}(t_0)|$; the inset: the amplitudes and the phases of the elements of the vector (\ref{V}). 
%$~_R\langle N_R|V|k\rangle_R$.
} 
  \label{Fig:P1ExtR10} 
\end{figure*}

In Fig.\ref{Fig:P1ExtR10}b, we  represent the amplitudes and phases of the elements of the column-vector (\ref{U2})  which appear in eq.(\ref{fN2})
and therefore they are the   elements of the matix $U$ which are most relevant to the optimization protocol. 
The profile  of the absolute values of the  projections $|f_n(t_0)|$ is shown
 in Fig. \ref{Fig:P1ExtR10}c.

For the fixed $N_S$, the critical length $N_c$  increases with an increase in $N_R$ and 
the  shapes of all curves shown in Fig.\ref{Fig:P1ExtR1}  change significantly. 
For instance, we represent 
the same characteristics for the case of $N_S=N_R=10$ in Fig.\ref{Fig:P1ExtR10}. In this case, the critical length of the communication line 
increases up to  $N_c=776$.
 Phase restriction (\ref{PhiInt}) (the horizontal dotted lines in Fig.\ref{Fig:P1ExtR10}a) selects the spectral interval 
 $457 \le k \le 574$ (the vertical dotted lines in Fig.\ref{Fig:P1ExtR10}a). Thus,  all valuable harmonics contribute  considerably to the 
 created projection $|f_N|$ (see also the inset in Fig.\ref{Fig:P1ExtR10}a). 
 
The amplitudes and phases of the   column-vector (\ref{U2}) and of  the row-vector (\ref{V})
are shown, respectively, in Figs.\ref{Fig:P1ExtR10}b and \ref{Fig:P1ExtR10}c (inset). These figures  confirm the symmetry (\ref{SVDsym22}).
In Fig.\ref{Fig:P1ExtR10}c, we also show the profile of the absolute values of the projections $|f_k(t_0)|$. 
We emphasize that the projections $f_n$, $n=N-N_R+1=767,\dots, N-1=775$ are identical to zero
which follows from the properties (\ref{U}) of the unitary transformations $U$ and $V$ in formula (\ref{fN2}).

In addition, we shall note that  the profile of the vector (\ref{U2}) in Fig.\ref{Fig:P1ExtR10}b is nothing but the initial profile of projections $f_n(0)$. In certain sense, it is  similar to the profile 
 $|\Psi_{N_S}\rangle =\frac{1}{\sqrt{N_S}}\sum_{m=0}^{N_S-1}(-1)^m|2 m+1\rangle$ found in Ref.\cite{BOWB} as the initial state 
 transferable along the spin-1/2 chain with high fidelity. But we have the  quasiperiodicity with the period of three spins in our case. 
\begin{table}
\begin{tabular}{|c|cccccccccc|}
\hline
\backslashbox[0.3cm]{{\small{$N_R$}}}{{\small{$N_S$}}}
              &1 &2        & 3       & 4     & 5      & 6     &  7    & 8      & 9     & 10 \\
\hline
1  &4 &4 &9 &11 &14&18&21&26&27&31\\
2  &4 &17&17&21 &27&45&47&52&56&71            \\
 3 &9 &17&22&26 &35&51&51&56&64&  83           \\
 4 &11&21&26&70 &83&93&102&126&172&   172         \\
 5 &14&27&35&83 &134&135&169&191&230&237\\
 6 &18&45&51&93 &135&139&180&196&235&239\\
 7 &21&47&51&102&169&180&293&340&353&407\\
 8 &26&52&56&126&191&196&340&449&452&547\\
 9 &27&56&64&172&230&235&353&452&458&564\\
 10&31&71&83&172&237&239&407&547&564&776\\  
 \hline
\end{tabular}
\caption{
The critical length  $N_c$ of the communication line performing the  high-probability state creation in dependence on the   
 dimensionalities of the sender $N_S$ and receiver $N_R$. There is a symmetry with respect to the exchange $R\leftrightarrow S$. 
} 
\label{Table:EVER} \label{Table:NSNRaHPST}
\end{table}

\subsection{Creating maximally mixed  states}
\label{Section:mixed}
It was shown in \cite{BZ_2015} that the remote creation of the eigenvalues of a quantum state is of principal importance   
because they cannot be modified by the 
local unitary transformation of the receiver, unlike the other parameters of the receiver state. 
%It was also shown that  the creating 
%the maximally mixed state (see  condition (\ref{mixed})) guaranties creating 
%the arbitrary eigenvalue  of the one-qubit receiver state  just varying the control parameters. 
The critical length for the maximally mixed state   created using the homogeneous communication line with $N_S=2$ and $N_R=1$ is 
$N_c=34$ nodes. But involving  the unitary transformations of  the two-qubit extended 
receiver (i.e., $N_R=2$) this length can be increased  up to $N_c=109$  nodes (see ref.\cite{BZ_2016}). 
In this paper we show that using the  proper initial state of the $N_S$-dimensional sender and the unitary transformation of the $N_R$-dimensional extended receiver 
(the matrices $U$ and $V$ in eq.(\ref{fN2}))
we can increase $N_c$ 
up to $N_c=164$ for the ten-qubit sender (and $N_R=1$). Moreover, increasing the dimensionality 
of the extended receiver we reach even better result: $N_c=5473$ for $N_S=N_R=10$.

\begin{figure*}
\subfloat[]{\includegraphics[scale=0.4,angle=0]{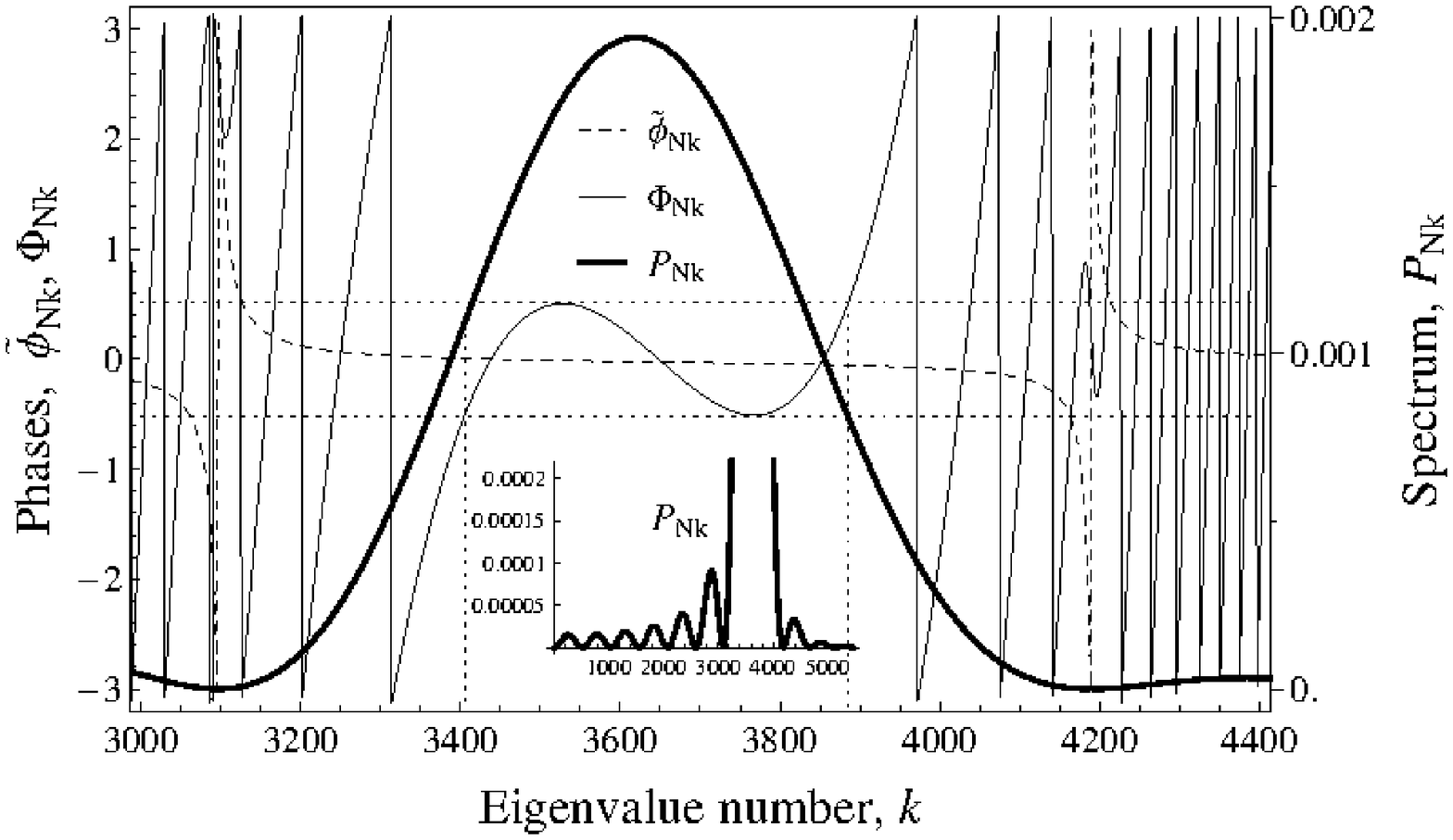}}
\subfloat[]{\includegraphics[scale=0.4,angle=0]{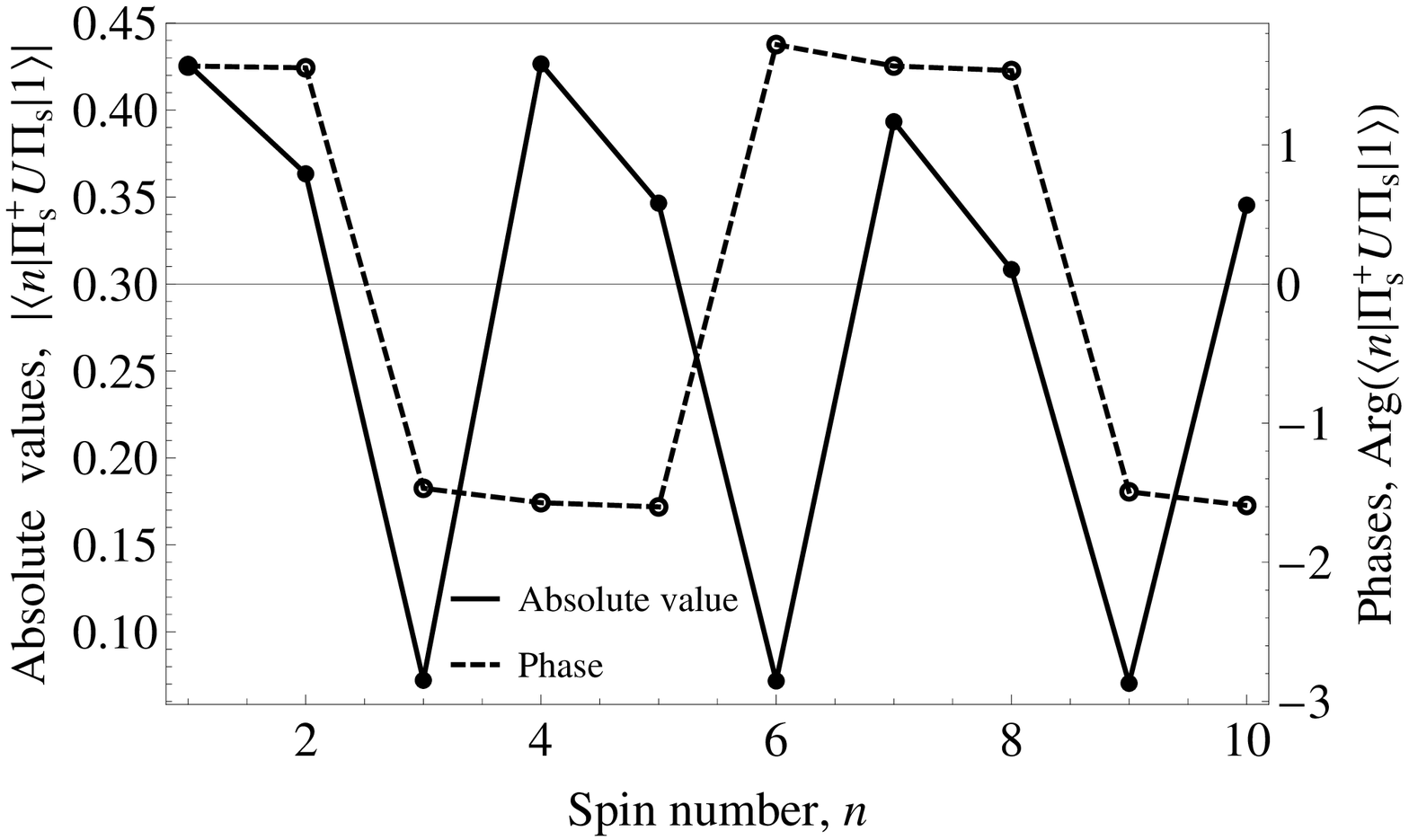}}\\
\hspace*{-0.6cm}\subfloat[]{\includegraphics[scale=0.49,angle=0]{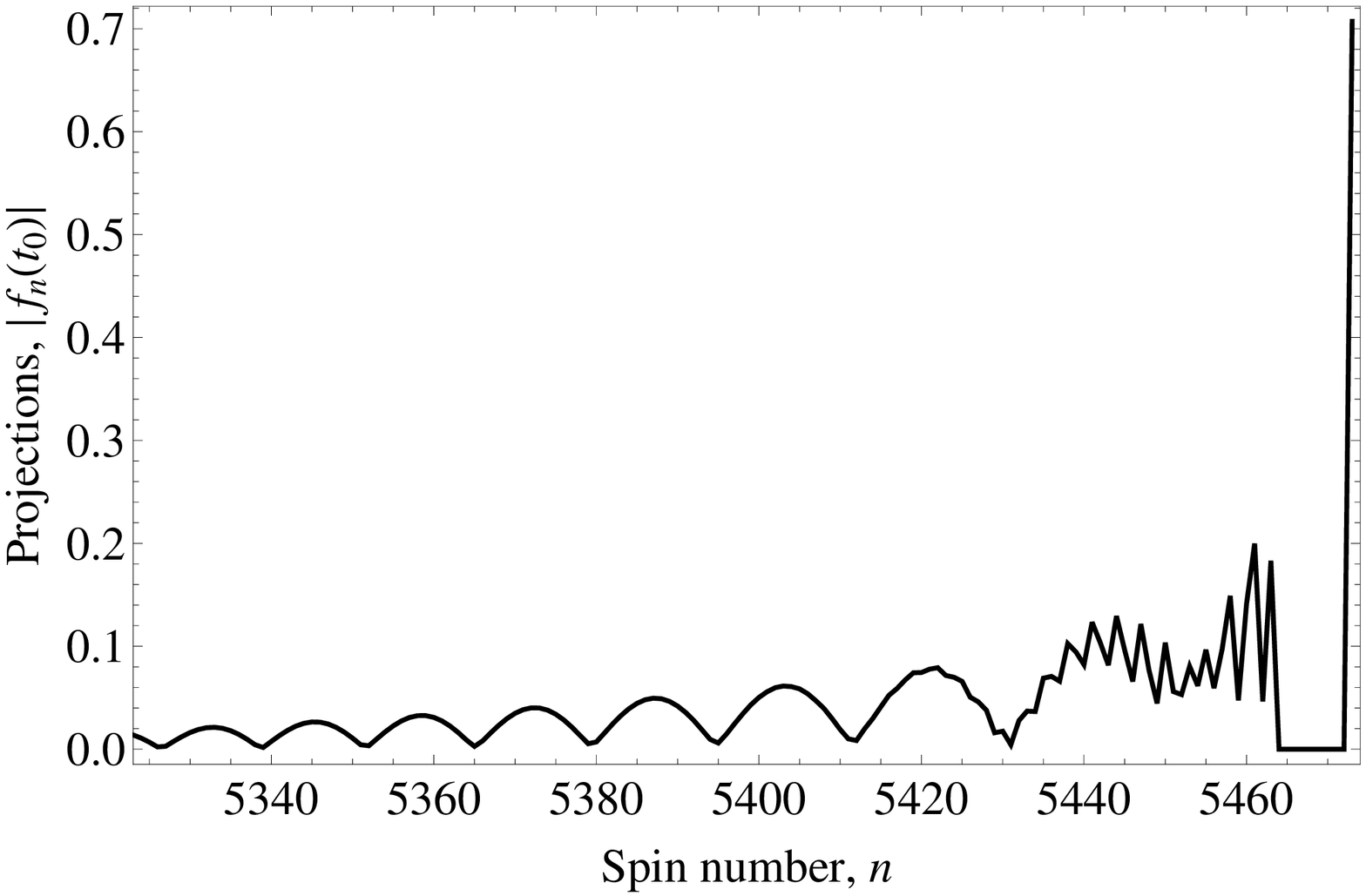}}

\vspace*{-8.1cm}

\subfloat{\includegraphics[scale=0.4,angle=0]{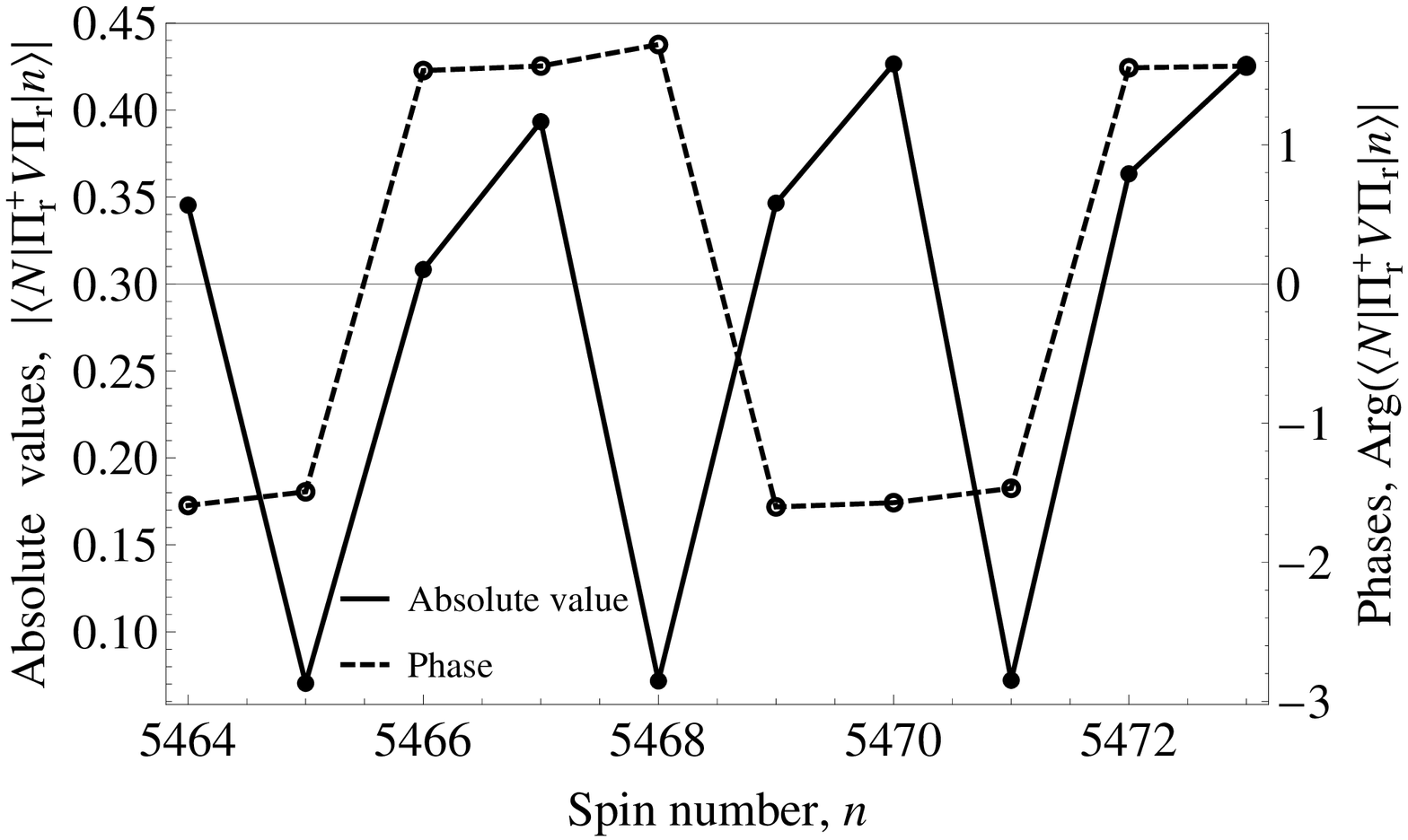}}

\vspace*{2cm}

\caption{%$R\sim 0.5$ n 5473 rsize 10, t=5404.02
The maximally mixed state (\ref{mixed})
in communication line of $N_c=5473$ nodes with   $N_S=N_R=10$  at $t_0=5404.02$.
(a) The spectrum $P_{Nk}$, the optimizing phases $\tilde \phi_{Nk}=\phi_{Nk} + \pi k$ and the  
resulting phases $ \Phi_{Nk}= \phi_{Nk}  - \lambda^H_{k} t_0$. The inset represent the spectrum  $P_{Nk}$ over the whole spectral interval;
(b) The  amplitudes and the phases of the elements of the vector (\ref{U2}) (or the initial projections 
$f_{n}(0)=\langle n|\Pi_S^+ U \Pi_S|1\rangle$);
(c) The profile of  the absolute values of the 
projections  $|f_{n}(t_0)|$; the inset: the amplitudes and the phases of the elements of the vector (\ref{V}).
%the amplitudes and phases of the profile $~_R\langle N_R|V|k\rangle_R$.
} 
\label{Fig:P05ExtR10} 
\end{figure*}

The critical length $N_c$ for different  dimensionalities $N_S$ and $N_R$  is given in  
Table \ref{Table:NSNRaLambda}.
\begin{table}
\begin{tabular}{|c|cccccccccc|}
\hline
\backslashbox[0.3cm]{{\small{$N_R$}}}{{\small{$N_S$}}}
             &1 &2        & 3       & 4     & 5      & 6     &  7    & 8      & 9     & 10 \\
\hline
1 & 22 & 37 & 45&63  &69  &106  &106  & 129&145&164\\
2 & 37 & 109&110&191 &256 &257  &294  &320&410&422\\
3 & 45 & 110&115&207 &265 &268  &314  &335&424&434\\
4 & 63 & 191&207&459 & 620&639  &876  &1000&1018&1183\\
5 & 69 & 256&265&620 &927 &937  &1316 & 1609&1616&1936\\
6 &106 & 257&268&639 &937 &948  &1344 &1628&1639&1977\\
7 &106 & 294&314&876 &1316&1344 &2031 &2490&2524&3199\\
8 & 129& 320&335&1000&1609&1628 &2490 & 3178&3198&4086\\
9 &145 &410 &424&1018&1616&1639 &2524 &3198&3223&4137\\
10 &164& 422&434&1183&1936& 1977&3199 &4086&4137&5473\\  
 \hline
\end{tabular}
\caption{
The critical length $N_c$ for the maximally mixed state 
 in dependence on the   
 dimensionalities of the sender $N_S$ and the receiver $N_R$. There is a symmetry with respect to the exchange $R\leftrightarrow S$. 
} 
   \label{Table:EVER2} \label{Table:NSNRaLambda}
   \end{table}
For the case  $N_c=5473$ and $N_S=N_R=10$, the spectrum $P_{Nk}$,  
the  phases $\tilde \phi_{Nk}=\phi_{Nk} +\pi k$ and  the 
resulting phases $\Phi_{Nk}= \phi_{Nk}  - \lambda^H_{k} t_0$ 
 are shown in Fig.\ref{Fig:P05ExtR10}a.
Condition (\ref{PhiInt}) (the horizontal dotted lines in Fig.\ref{Fig:P05ExtR10}a) 
selects the spectral interval $3405 \le k \le 3886$  (the vertical dotted lines). Therefore,   
unlike the case of high-probability state creation, not all large-amplitude harmonics are involved in the state-creation process
(see also the inset in Fig.\ref{Fig:P05ExtR10}a).

Finally, we represent the amplitudes and phases of the elements of the column vector (\ref{U2}) and the row-vector (\ref{V}), 
respectively,  in Figs.\ref{Fig:P05ExtR10}b and \ref{Fig:P05ExtR10}c (inset). Similar to Fig. \ref{Fig:P1ExtR10}b,c these figures  confirm the symmetry (\ref{SVDsym22}). In Fig.\ref{Fig:P05ExtR10}c we also show the profile of the absolute values of the projections $|f_n(t_0)|$.
Here, $f_n$, $n=N-N_R+1=5464,\dots, N-1=5472$ are identical to zero because of  the  unitary transformations $U$ and $V$ in formula  (\ref{fN2}).

\section{Two-qubit state creation}
\label{Section:2qubit}

In order to cover the  large region of the two-qubit receivers state  space we need to replace the one-excitation initial state (\ref{InitialSt}) with the  two-excitation  one:
\begin{eqnarray}\label{InSt2}
|\Psi_0\rangle=%a_0 |0\rangle + |\psi_0\rangle,\;\;|\psi_0\rangle = 
a_0 |0\rangle+\sum_{i=1}^{N_S} a_i |i\rangle + \sum_{{i,j=1}\atop{j>i}}^{N_S} a_{ij} |ij\rangle,\;\;\sum_{i=0}^N |a_i|^2 +
\sum_{{i,j=1}\atop{j>i}}^{N_S} |a_{ij}|^2=1, 
\end{eqnarray}
{where $|ij\rangle$ is the state with the $i$th and $j$th spins excited. }
Now, instead of (\ref{RhoR}), the density matrix $\rho^R$ is given by the formula
\begin{eqnarray}\label{RhoR2}
\rho^R = {\mbox{Tr}}_{1,\dots,N-2} |\Psi^V\rangle \langle \Psi^V|,\;\;|\Psi^V\rangle = e^{-i H t}|\Psi_0 \rangle ,
\end{eqnarray}
where trace is taken over all the nodes except the nodes of the receiver.
Before proceeding to the state creation we need to determine the time instant $t_0$
for the state registration. We take the time instant maximizing the quantity 
\begin{eqnarray}\label{varkappa}
\varkappa=\langle N|\rho^R|N\rangle + \langle N-1|\rho^R|N-1\rangle + \langle N,N-1|\rho^R|N,N-1\rangle 
\end{eqnarray}
averaged over the initial states (\ref{InSt2}), 
\begin{eqnarray}\label{max}
\max_t \langle \varkappa \rangle =\langle \varkappa \rangle|_{t=t_0},
\end{eqnarray}
where $\langle \cdot \rangle$ means the average over pure initial states of the sender. 
The formula simplifying the calculation of $t_0$ is derived in 
Appendix, Sec. \ref{Section:AppendixA}.

\begin{figure*}
\includegraphics[scale=0.5,angle=0]{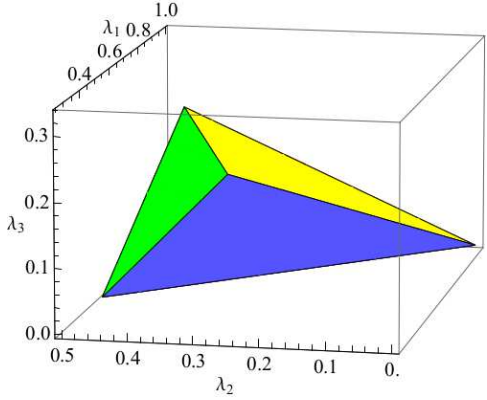}
\caption{ The  three-dimensional region  of the independent spectral parameters  $\lambda_i$, $i=1,2,3$ of the two-qubit receiver's state space is represented by the tetrahedron
with the vertexes $L_1=(1,0,0)$, $L_2=\frac{1}{2}(1,1,0)$, $L_3=\frac{1}{3}(1,1,1)$ and  $L_4=\frac{1}{4}(1,1,1)$. The whole  tetrahedron can be created only in the short communication lines with $N\le 16$. 
} 
  \label{Fig:tetr} 
\end{figure*}

The creation of two-qubit states requires using the two-excitation dynamics and therefore this case becomes more  complicated for numerical simulations. 
We restrict ourselves to the four-qubit sender ($N_S=4$) and two-qubit receiver 
without involving  the extended receiver ($N_R=2$), and consider only the eigenvalue creation.  The all possible values of three independent eigenvalues of the two-qubit state 
form a  tetrahedron in the three-dimensional space of independent   eigenvalues $\lambda_i$, $i=1,2,3$ (remember that $\lambda_4=1-\sum_{i=1}^3\lambda_i$)
with the vertexes $L_1=(1,0,0)$, $L_2=\frac{1}{2}(1,1,0)$, $L_3=\frac{1}{3}(1,1,1)$ and  $L_4=\frac{1}{4}(1,1,1)$, corresponding to the states, respectively, 
 $\rho^R_{1}={\mbox{diag}} (1,0,0,0)$, $\rho^R_{2}=\frac{1}{2} {\mbox{diag}} (1,1,0,0)$,
$\rho^R_{3}=\frac{1}{3} {\mbox{diag}} (1,1,1,0)$  
and
$\rho^R_{4}=\frac{1}{4}{\mbox{diag}} (1,1,1,1)$, as  shown in Fig.\ref{Fig:tetr}.

Of course, not all points inside of this tetrahedron are achievable in long chains, so that each point $(\lambda_1,\lambda_2,\lambda_3)$ has its own critical length $N_c(\lambda_1,\lambda_2,\lambda_3)$ such that this point is not creatable in the communication line with $N>N_c$.
For instance, the vertex $L_1$ is creatable in the communication line of any length, i.e.  $N_c(L_1)=\infty$. In addition, 
$N_c(L_2)=191$ was calculated in Sec.\ref{Section:mixed}\footnote{Although $N_c(L_2)$ was calculated in  Sec.\ref{Section:mixed} for the case of one-excitation dynamics, 
it is verified  numerically that $N_c(L_2)$ remains the same if we involve the two-excitation dynamics}. The direct calculations (see Appendix,
Sec. \ref{Section:AppendixB}, for details) show that 
$N_{c}(L_3)=N_{c}(L_4)=16$. Thus, the whole tetrahedron can be created only in the short communication lines with $N\le 16$.

\begin{figure*}
\subfloat[]{\includegraphics[scale=0.8,angle=0]{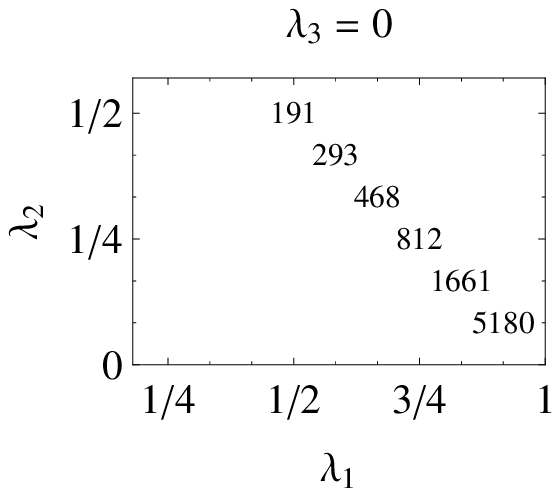}}
\subfloat[]{\includegraphics[scale=0.8,angle=0]{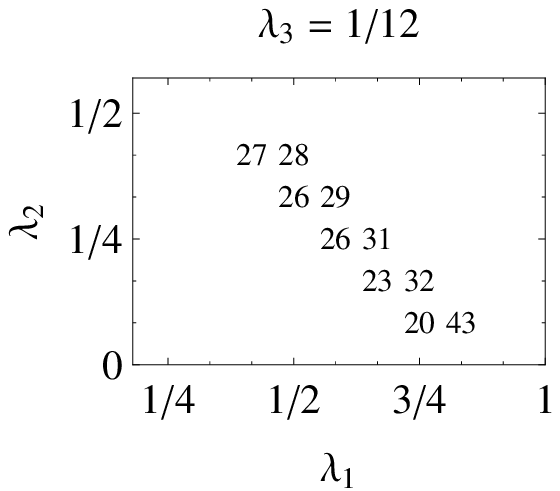}}\\
\subfloat[]{\includegraphics[scale=0.8,angle=0]{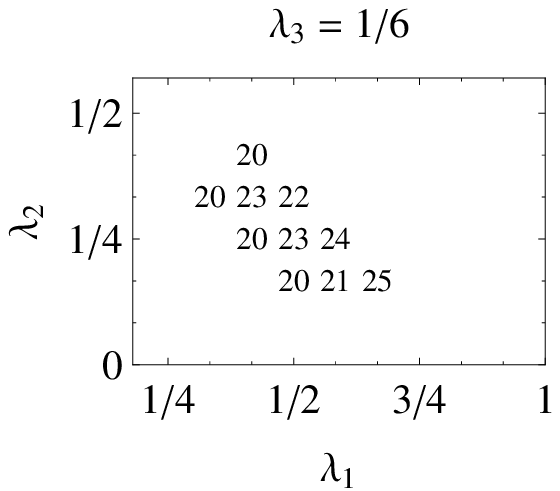}}
\subfloat[]{\includegraphics[scale=0.8,angle=0]{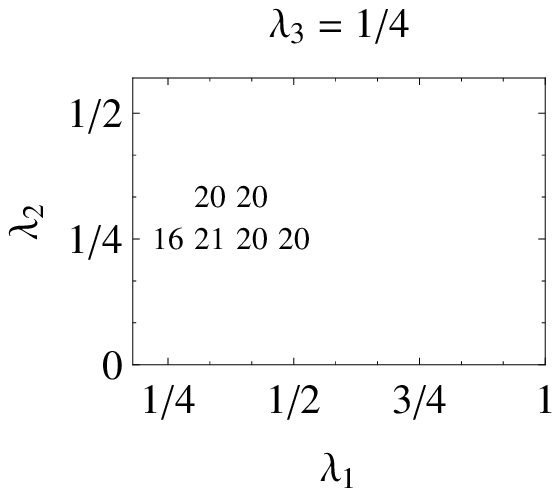}}
\caption{The critical length  $N_c$ in dependence on the eigenvalues of the creatable state for the lattice of points (\ref{lattice}).  
  $N_c$ is indicated  on the plane $(\lambda_1,\lambda_2)$ by the integer numbers,  while $\lambda_3$ is given as a parameter in  figures (a)-(d). 
 Two critical lengths are  missing. The first one  corresponds to 
  $\lambda_3=\frac{1}{3}$: $N_c(\frac{1}{3},  \frac{1}{3},  \frac{1}{3})=16$.
 The second critical length $N_c(1,0,0)=\infty$ must be at the right lower corner of figure (a). 
The critical length increases with approaching the line $\lambda_3=0$, 
$\lambda_2=1-\lambda_1$. The figure (a) is constructed using one-excitation initial state and 
the optimization protocol based on the SVD. 
The optimization in the case $\lambda_3>0$, figures (b)-(d), is performed using the approach of Sec.\ref{Section:2qubit}.
} 
  \label{Fig:lattice} 
\end{figure*}

 To get some overview of  critical lengths $N_c$ as a function of eigenvalues  we calculate them  for the  lattice  of points 
\begin{eqnarray}\label{lattice}
 \lambda_1\ge  \lambda_2\ge  \lambda_3,\;\; 
 \lambda_i =\frac{p_i}{12}, \;\;p_i\in \ZZ
 \end{eqnarray} 
%with the interval   
%$|\lambda_j^{(n_k+1)}- \lambda_j^{(n_k)}| =\frac{1}{12}$, $k,j=1,2,3$,  
as shown in Fig.\ref{Fig:lattice}.
We can conclude  that, with an increase in $N$, the creatable points are accumulating around  the edge of the tetrahedron
connecting the vertexes $L_1$ and $L_2$, see Fig.\ref{Fig:lattice}a. Notice that calculating $N_c$ for the case $\lambda_3=\lambda_4=0$ 
shown in Fig.\ref{Fig:lattice}a we use the one-excitation initial state and 
the optimization protocol based on the SVD (see Secs.\ref{Section:SVD}, \ref{Section:1qubit}).  The two-excitation initial state 
does not change the result,  which is confirmed by the numerical simulations.

\section{Conclusion}

\label{Section:conclusion}
In this paper we consider the optimization problem of  the remote creation of quantum states . %In the case of one-qubit receiver, we consider 
%the creation of almost pure and maximally mixed states. In thecase of two-qubit receiver, we restrict ourselves to the problem of eigenvalue %creation.   
As  the optimization tool we  use
the  unitary transformations  of the $N_S$-qubit sender and $N_R$-qubit 
extended receiver (which includes the receiver as a subsystem). 
For the one-qubit state, we consider the creation of two types of states: (i) (almost) pure state with single excitation of one-qubit receiver (this process is an analogue of the HPST)
and (ii) the creation of maximally mixed state (the state with two equal eigenvalues). In both cases we investigate the dependence of the critical length on the dimensionality of the sender ($N_S$) and extended receiver ($N_R$)
showing that $N_c$  grows with an increas in  both $N_S$ and $N_R$. Thus, restricting ourselves to the 10-qubit sender and extended receiver we achieve 
$N_c=776$ for the high probability state creation and $N_c=5473$ for the creation of the maximally mixed state. 

Considering the spectral representation of the state evolution we explicitly demonstrate that the optimal parameters of the sender's initial state and of the unitary transformation of the extended receiver 
provide (i) the large  spectral amplitudes and (ii) the proper spectral phases. Thus, almost the whole spectrum with large spectral amplitudes is involved into the high-probability state creation process (i.e., the spread of their phases satisfies  condition (\ref{PhiInt})), whereas only half of them serve to create the maximally mixed state.

The creation of two-qubit states is more complicated because it is based on  the two-excitation dynamics. Therefore we study only a particular example of communication line with   four-qubit sender and two-qubit receiver (without the extended receiver) and consider the problem of eigenvalue creation. The creatable subregion form the tetrahedron in the three-dimensional space of the independent eigenvalues $\lambda_i$, $i=1,2,3$.
We study the critical length $N_c$ as a function of a point inside of the above tetrahedron    (i.e., the maximal  length of the communication 
line allowing us to create a particular point inside of the tetrahedron).
In particular, we have found $N_c(L_1)=\infty$, $N_c(L_2)=191$,  $N_{c}(L_3)=N_{c}(L_4)=16$.
Therefore, any point inside of the tetrahedron can be created using 
the communication line of $N\le 16$ nodes.

This work is partially supported by the program of RAS 
''Element base of quantum computers''(No. 0089-2015-0220) and 
by the Russian Foundation for Basic Research, grant No.15-07-07928.

\section{Appendix}
\label{Section:Appendix} 
\subsection{Optimal time instant for state registration}
\label{Section:AppendixA} 
We find the time instant  maximizing the quantity $\varkappa$ (\ref{varkappa})
averaged over the initial states (\ref{InSt2}), see Eq.(\ref{max}).
By its definition (\ref{RhoR}), $\rho^R$  is a Hermitian form  (the exact formula for $\rho^R$  is given in \cite{SZ_2016}). Consequently, 
the quantity $\varkappa$ is a Hermitian form as well and can be written as 
\begin{eqnarray}
\varkappa(|\Psi_0\rangle )=\langle \Psi_0|\Pi_S^+ A \Pi_S|\Psi_0\rangle,
\end{eqnarray}
where $A$ is a Hermitian operator independent on the  parameters $a_i$ and $a_{ij}$.
We can diagonalize  $A$:
\begin{eqnarray}
A={\cal{U}} \Omega {\cal{U}}^+,
\end{eqnarray}
where $\Omega ={\mbox{diag}}(\tilde \lambda_1\dots \tilde \lambda_{\tilde N_S}) $
is the diagonal matrix of the eigenvalues and ${\cal{U}}$ is the corresponding matrix of the eigevectors of $A$. 
Here, $\tilde N_S=N_S(N_S+1)/2 +1$ is the dimensionality of the 
basis of the sender's state  space, whose elements have the form (\ref{InSt2}). 
Obviously,
\begin{eqnarray}\label{Avr}
&&
\langle \varkappa(\Psi_0)\rangle =\big\langle \langle \Psi_0|\Pi_S^+ A \Pi_S|\Psi_0\rangle \big\rangle =
\big\langle \langle \Psi_0|\Pi_S^+ \Omega \Pi_S|\Psi_0\rangle \big\rangle =\\\nonumber
&&
\sum_{j=1}^{\tilde N_S} \big\langle \langle \Psi_0|\Pi_S^+{\mbox{diag}}(\tilde \lambda_j \underbrace{0\dots 0}_{\tilde N_S-1}) \Pi_S|\Psi_0\rangle \big\rangle =\frac{1}{\tilde N_S}
\sum_{j=1}^{\tilde N_S} \big\langle \langle \Psi_0|\Pi_S^+{\mbox{diag}}(\underbrace{\tilde \lambda_j \dots \tilde \lambda_j}_{\tilde N_S}) \Pi_S|\Psi_0\rangle \big\rangle 
=\\\nonumber
&&\sum_{j=1}^{\tilde N_S} \frac{\tilde \lambda_j}{\tilde N_S}=\frac{{\mbox{Tr}} A}{\tilde N_S} ,
\end{eqnarray}
In turn,
\begin{eqnarray}\label{Avr2}
{\mbox{Tr}} A = \sum_{i=1}^{N_S} \langle i | \Pi_S^+ A \Pi_S |i\rangle + \sum_{{i,j=1}\atop{j>i}}^{N_S} \langle i j | \Pi_S^+ A \Pi_S |i j\rangle=
\sum_{i=1}^{N_S} \varkappa(|i\rangle) + \sum_{{i,j=1}\atop{j>i}}^{N_S} \varkappa(|ij\rangle).
\end{eqnarray}
All in all, Eqs.(\ref{Avr}) and (\ref{Avr2}) yield 
\begin{eqnarray}\label{Avr3}
&&
\langle \varkappa(\Psi_0)\rangle = \frac{1}{\tilde N_S} \left(
\sum_{i=1}^{N_S} \varkappa(|i\rangle) + \sum_{{i,j=1}\atop{j>i}}^{N_S} \varkappa(|ij\rangle)\right).
\end{eqnarray}
Consequently, we need to find the time instant $t_0$
maximizing  the rhs of Eq.(\ref{Avr3}). Thus  the multi-parameter optimization is reduced  to the optimization over the single parameter $t$.

\subsection{Eigenvalue creation in two-qubit receiver}
\label{Section:AppendixB}

The parameters $a_i$ and $a_{ij}$ in the initial state (\ref{InSt2})  which  result to the states with four and three equal eigenvalues (the vertex $L_3$ and $L_4$ of 
the tetrahedron in Fig.\ref{Fig:tetr}) can be found 
using  the characteristic equation 
\begin{eqnarray}\label{char}
|\rho^R -\lambda I|= \lambda^4 - \lambda^3 + A \lambda^2 + B \lambda +C,\;\;\;C=\det  \rho^R,
\end{eqnarray}
where $A$, $B$ and $C$ are functions of the parameters $a=(a_i:i=1,\dots,N, a_{ij}:, j>i, i=1,\dots,N-1,\;j=2,\dots,N)$ and time $t$.
In general, if we are interested in a state with the fixed set of eigenvalues $\lambda_i$, $i=1,2,3$, then 
the coefficients of the  characteristic equation are the known functions  $A_0$, $B_0$ and $C_0$ of $\lambda_i$, 
and we have to find the set of  parameters $a$ solving the following  system at the optimized time instant $t_0$ (found in  Appendix, Sec.\ref{Section:AppendixA}): 
\begin{eqnarray}
\label{char2} 
A(a,t_0)=A_0(\lambda_1,\lambda_2,\lambda_3),\;\;B(a,t_0)=B_0(\lambda_1,\lambda_2,\lambda_3),\;\;C(a,t_0)=C_0(\lambda_1,\lambda_2,\lambda_3).
\end{eqnarray}
To find the  approximate solution to this system we minimize the discrepancy 
\begin{eqnarray}
\varepsilon(\lambda_1,\lambda_2,\lambda_3) = \sqrt{ \left(A-A_0\right)^2+\left(B-B_0\right)^2+\left(C-C_0\right)^2}.
\end{eqnarray}
If the minimized values of $\varepsilon$ exceeds the certain value  for some $N$, then we say that the state is not creatable in the communication line of length $N$.

\begin{figure*}
\includegraphics[scale=0.8,angle=0]{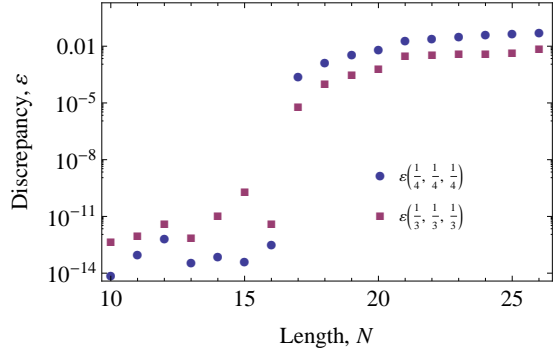}
\caption{ The accuracies $\varepsilon(L_3)$ and  $\varepsilon(L_4)$
as  functions of the length of communication line.
} 
  \label{Fig:accuracy} 
\end{figure*}

To clarify the above arguments, we consider the creation of the vertexes  $L_3$ and $L_4$ of the tetrahedron in Fig.\ref{Fig:tetr}. For them  
we have, respectively,
$A_0(\frac{1}{3},\frac{1}{3},\frac{1}{3}) =\frac{1}{3}$,
$B_0(\frac{1}{3},\frac{1}{3},\frac{1}{3}) = -\frac{1}{27}$, $C_0(\frac{1}{3},\frac{1}{3},\frac{1}{3})=0$
and $
A_0(\frac{1}{4},\frac{1}{4},\frac{1}{4}) =\frac{3}{8}$, $B_0(\frac{1}{4},\frac{1}{4},\frac{1}{4}) =-\frac{1}{16}$,
$C_0(\frac{1}{4},\frac{1}{4},\frac{1}{4}) =\frac{1}{256}$.
The accuracies  $\varepsilon(L_3)$ and  $\varepsilon(L_4)$ as  functions of $N$ are shown 
in Fig.\ref{Fig:accuracy}. Passing from $N=16$ to $N=17$, both $\varepsilon(L_3)$ and  $\varepsilon(L_4)$ jump, respectively, 
from $\sim 10^{-12}$ to $\sim 10^{-5}$ and from $\sim 10^{-13}$ to $\sim 10^{-4}$. 
This indicates that the vertexes 
$L_3$ and $L_4$ are not achievable for $N>16$.  
Therefore we conclude that  $N_c(L_4)=N_c(L_3)=16$.

\end{document}